\documentclass[12pt,a4paper]{article}
\usepackage{a4wide}
\usepackage{aas_macros}
\usepackage[utf8]{inputenc}
\usepackage[T1]{fontenc}
\usepackage{hyperref}
\usepackage{latexsym,amsbsy,amsmath,bm}
\usepackage[pdftex]{graphicx}
\usepackage{xcolor}
\usepackage{longtable}
\usepackage{isotope}
\usepackage{microtype}
\usepackage{cite}
\usepackage[bitstream-charter]{mathdesign}
 \DeclareGraphicsExtensions{.eps, .jpg,.png}
\renewcommand{\vec}[1]{\boldsymbol{#1}}

\def\ll#1#2{\tilde{\lambda}_{#1}.\tilde{\lambda}_{#2}}
\def\llss#1#2{\tilde{\lambda}_{#1}.\tilde{\lambda}_{#2}\,\boldsymbol{\sigma}_{#1}\boldsymbol{\sigma}_{#2}}

\makeatletter
\newcommand*{\rom}[1]{\expandafter\@slowromancap\romannumeral #1@}
\makeatother
\usepackage{palatino}
\begin{document}
\title{Hadrons and Few-Body Physics}
%
%
\author{Jean-Marc~Richard\\
{\small \texttt{j-m.richard@ipnl.in2p3.fr}}\\
{\small Universit\'e de Lyon, Institut de Physique des 2 Infinis de Lyon,
IN2P3-CNRS--UCBL}\\
{\small 4 rue Enrico Fermi, 69622 Villeurbanne, France}}
\date{\today}
\maketitle
\begin{abstract}
We present a selection of topics with an interplay of hadron and few-body physics. This includes  few-nucleon systems, light hypernuclei and quark dynamics for baryons and multiquarks. It is stressed that standard quark models predict very few stable multiquarks.
\end{abstract}
\section{Introduction}
\label{se:intro}
There are several instances where  few-body physics interplays with hadron dynamics. Obviously  the properties of light nuclei and the scattering of proton or deuteron on light nuclei constitute the first examples. Its natural extension towards light hypernuclei is also particularly interesting as some systems are at the edge of stability. With the advent of the quark model in the early 60s, few-body techniques have been applied to three-quark and quark-antiquark systems, and more recently to higher configurations, in the search for stable or metastable multiquark systems. 
\section{Few-nucleon systems}
\label{se:few-nucleon}
Various methods  have been developed during the years to study the light nuclei with the successive versions of the  nucleon-nucleon potential. Among them, one can cite the hyperspherical expansion~\cite{Marcucci:2019hml} or the solution of Faddeev-type of equations~\cite{Lazauskas:2020qzo}.
The corresponding techniques have proved  useful for other systems.

Even the famous Efimov effect~\cite{Efimov:1970zz} has been inspired by few-nucleon systems, with the observation that the nucleon-nucleon interaction is at the edge between binding and non-binding, on one side for neutron-proton with isospin $I=0$ and spin $I=1$, and on the other side for proton-proton with isospin $I=1$ and spin $S=0$.  

The ancestor of the Efimov effect\footnote{When receiving the Faddeev medal at the Few-Body Conference held in Caen \cite{Orr:2020uml}, Efimov gave a very interesting account on how he arrived to this new effect by studying in depth the paper by Thomas}, the Thomas collapse~\cite{Thomas:1935zz}, comes from a comparison of the three-body ground-state energy $E_3$ and the two-body  $E_2$. From the data available on $E_3/E_2$, Thomas inferred that the meson predicted by Yukawa but not yet discovered at that time, should be lighter than 200\,MeV. 

More recently, the study of three-body systems  \isotope[3]{H} and \isotope[3]{He} and other light nuclei using $NN$ potentials that fit the deuteron energy $E_2$ and the $NN$ phase-shifts, has revealed the need for three-body forces.

This domain is still very active, with, for instance, the speculations about the dineutron and tetraneutron~\cite{Deltuva:2019mnv}. 
\section{Light hypernuclei}
Since the early 50s, there is a continuous activity on hypernuclei with many interesting questions such as the modification of the weak decay for bound $\Lambda$ hyperons. Most of the identified hypernuclei have strangeness $S=-1$, but states with $S=-2$ have also been found. For years, progress on the knowledge of the $\Lambda N$ and $\Lambda\Lambda$ interactions has been purely theoretical, and benefited from our improved approach to the nucleon-nucleon interaction (meson-exchanges, SU(3) constraints, effective theories). There is a reasonable hope, however, to get information on baryon-baryon systems from the correlations of their production through heavy-ion collisions~\cite{Acharya:2020asf}. The successive versions of the Nijmegen potential \cite{Rijken:2019llc} and the interactions derived from the chiral effective theories \cite{Haidenbauer:2019boi} have motivated successive upgrades of the studies of the strange few-baryon systems. For a review, see, e.g., \cite{Hiyama:2012gx,Contessi:2019csf,Haidenbauer:2019boi,Garcilazo:2020edy}.  It was rediscovered in this context that while 2-body binding is directly related to the sign of the scattering length, the possibility of Borromean $n$-body binding with $n\ge3$ depends crucially on the magnitude of the effective range~$r_e$. For instance the penultimate version of the $\Lambda\Lambda$ interaction from the chiral effective theories had a very small $r_e$, which allowed the possibility of binding $\Lambda\Lambda n n$~\cite{Richard:2014pwa}, but the latest version \cite{Haidenbauer:2019boi} has a much larger $r_e$ which prohibits the binding of $\Lambda\Lambda n n$.

The physics of hypernuclei has been extended to charm and beauty. Any hadron containing at least a light quark or antiquark participates in some Yukawa-type of interaction. Some of $XYZ$ exotic mesons or LHCb pentaquarks have even been anticipated in this scheme. For a review on such ``molecules'', see, e.g., \cite{Brambilla:2019esw}. Perhaps this approach is too flexible, as dozens of states are predicted. Note also that many hadron-hadron pairs are found with a large scattering length. Thus some three-hadron systems are likely to be bound~\cite{Bicudo:2004pr,Huang:2019qmw}.

\section{Quark model of mesons}
Somewhat paradoxically, explicit quark model calculations were developed first for the case of baryons, as seen in the next section. But the situation changed in 1974 with the discovery of the $J/\psi$ and its excitations, and their interpretation as $c\bar c$ bound states. The Schrödinger equation was rediscovered by a community more accustomed to dispersion relations and Feynman diagrams. 
The charmonium is studied by solving 
\begin{equation}\label{eq:Schr}
 \left[-\frac{\Delta}{m}+V\right]\Psi=E\,\Psi~
\end{equation}
and matching the meson masses $2\,m+E$ with the experimental data, as well as some observables, such as the radiative transitions.  Here $m$ is the constituent mass of the charmed quark, and the potential is inferred as
\begin{equation}\label{eq:pot-cc}
 V= -\frac{a}{r}+ b\,r + c + \cdots
\end{equation}
where the ellipses denote the spin corrections. Among the rediscoveries, one might mention the astute observation~\cite{DiasdeDeus:1980vzk} that the variational solutions obey the same scaling laws and the same virial theorem as the exact solution, already pointed out (independently and in a different context) by Hylleraas and Fock around 1930 \cite{1929ZPhy...54..347H,1930ZPhy...63..855F}. The charmonium gave also the opportunity to extend some theorems on the level order, inverse problem, wave function at the origin, etc.~\cite{Quigg:1979vr,Grosse:1979xm,Grosse:847188}. 

For the phenomenology of charmonium, it was quickly realized that the potential such as \eqref{eq:pot-cc}, with four parameters $a$, $b$, $c$ and $m$, is far from being unique. The game became more interesting after the discovery of the $\Upsilon$ family, i.e., the $b\bar b$ spectrum, and even anticipated before: is it possible to reproduce the spin-averaged levels of charmonium and bottomonium with the \emph{same} central potential? This turns out to be possible, and this provides a test of flavor independence: the gluons that build the potential are coupled to the color of the quarks, not to their flavor. 
\section{Quark model of baryons}
As already mentioned, the first detailed quark-model calculations were devoted to baryons, with, in particular, the pioneering papers by Greenberg~\cite{Greenberg:1964pe} and Dalitz~\cite{Dalitz:1965fb}. If one takes the approximation of an  harmonic oscillator 
\begin{equation}\label{eq:HO1}
 V=K\left( r_{12}^2+r_{23}^2+r_{31}^2\right)~,
\end{equation}
one gets an intrinsic Hamiltonian (free of center-of-mass motion)
\begin{equation}\label{eq:HO2}
 H=\frac{\vec p_\rho^2}{m}+\frac{\vec p_\lambda^2}{m'}+ \frac{3\,K}{2}\left(\vec\rho^2+\vec\lambda^2\right)~,
\end{equation}
where $\vec\rho=\vec r_2-\vec r_1$ and $\vec\lambda=(2\,\vec r_3-\vec r_1-\vec r_2)/\sqrt3$ are the usual Jacobi coordinates, and $m'=m$ for an equal-mass baryon $mmm$ of the $N$ or $\Delta$ sector, and $m'=3\,m\,M/(2 M+m)$ for a $mmM$ hyperon. To account for the observation of  very moderate splittings in the SU(3) multiplets $N,\,\Lambda,\,\ldots$ and $\Delta,\,\Sigma^*\,\ldots$, one needs to use the ground state solution  of \eqref{eq:HO2}, that is symmetric under permutation. But, when a symmetric spatial wave function is used for identical quarks in the ground state, one cannot ensure the Fermi statistics. The outcome is well-known:  a ``para-statistics'' was first advocated for the quarks, and color was invented, that eventually led to Quantum Chromodynamics. 

The question is how to solve the three-quark problem for a more general interaction. Let us assume an Hamiltonian
\begin{equation}\label{eq:H-baryons}
 H=\sum_{i=1}^3\frac{\vec p_i^2}{2\,m_i}+\frac12\sum_{1\le i<j\le3} v(r_{ij})~,
\end{equation}
where, for instance, the potential is the popular  $v(r)=-a/r+b\,r+c$, and the 1/2 factor is discussed in the next section. The strategies used have been diverse. One could mention:
\begin{itemize}
 \item Keep using  the HO wave function and treat the anharmonicity as a first-order perturbation.  This is how Isgur, Karl and their collaborators convinced us that the quark model is far from being unfriendly, and actually can capture a lot of physics with minimal tools~\cite{Isgur:1980mb}. Their precursors were Dalitz et al.\ \cite{Dalitz:1965fb,Horgan:1973ww}, Hey et al.~\cite{Hey:1982aj}, and their (friendly) competitors such as Gromes and Stamatescu \cite{Gromes:1979xn} and many others.  
 \item For each state, use a basis with the lowest HO wave function and its excitations.  There are only two parameters: the oscillator strength and the size of the basis. However, the convergence is not very satisfactory for the wave function. 
 \item Use a basis of correlated Gaussians, either in a variant of Kamimura et al.~\cite{Hiyama:2003cu}, or of Suzuki and Varga~\cite{1999fbpp.conf...11V}. 
 \item Organize a hyperspherical expansion \cite{Giumaraes:1981gn,Richard:1992uk}
 \item Solve the Faddeev equations, that have been translated into position-space potentials and adapted to the confining interactions \cite{SilvestreBrac:1985ic}.
\end{itemize}

The Born-Oppenheimer (BO) approximation was devised in 1927, and is widely used in most areas of few-body physics. It took some time, unfortunately,  before the method was explicitly used in quark-model calculations. However, for a doubly-heavy baryon, this is an obvious choice. After removing the center of mass motion, one solves the one-body problem with two centers, corresponding to a reduced mass $3\,m\,M/(m+3\,M)$, and the energy, supplemented by the direct $QQ$ interaction, builds the effective $QQ$ interaction.\footnote{The naive BO consists in solving the two-center problem with the light-quark mass $m$. The center-of-mass energy is the main (and most obvious) correction to BO}\@
Actually, everything is BO in this field. The quarkonium potential is the energy of the gluons field for a given interquark separation. The second gluon energy is interpreted as the effective potential for hybrids, etc. See, e.g.~\cite{Braaten:2013boa} and refs.\ there. 

The diquark approximation (DQ) is just opposite to BO, in which the interquark separation $\bm R$ is first frozen, and the wave equation solved for $\bm r$, the distance of the light quark to the center of mass of the heavy quarks, and in a second step, the Schrödinger equation is solved for $R$. In the most naive version of the diquark scheme, one first solves for $\bm R$ using solely the direct $QQ$ interaction, and then one solves for $r$ assuming that the interaction between $q$ and $QQ$ is twice the $qQ$ interaction. Except for a pure Coulomb interaction (Gauss theorem), this results in an artificial lowering of the energies. For instance, for a harmonic potential $\sum r_{ij}^2$ and the standard choice of Jacobi coordinates, the naive diquark approximation corresponds to a factorization into $\bm R^2$ associated to the mass $M$ and $\bm r^2$ associated to  the reduced mass, while the exact solution involves $3\,{\bm R}^2/2$ for the former. This means a an error of a factor $\sqrt{2/3}$ in the leading term. Moreover, the first excitations are within the $\bm R$ degrees of freedom: in the DQ approach, one has to re-estimate the diquark for each level, while the BO gives simultaneously all levels of  the lower part of the spectrum.

Note that the concept of diquark is very useful \cite{Anselmino:1992vg}, and as it leads to a simplification in many instances. For instance, it was first \emph{assumed} that the excited baryons consist of a quark and a diquark, to explain that the slope of the Regge trajectory is the same for baryons and mesons. See, for instance, \cite{Eguchi:1975id}. It was later \emph{demonstrated} that within a large class of Hamiltonians, the excited baryons acquire spontaneously a quark-diquark structure \cite{Martin:1985hw}. It would be interesting to extend this study to four-quark states and see whether for high angular momentum $L$, the lowest $qq\bar q\bar q$ states consists of an $S$-wave diquark and an $S$-antidiquark whose relative motion carries the whole $L$, i.e., the phenomenological conjecture of baryonium~\cite{Jaffe:1977cv,Chan:1978nk}.

What is less convincing is the blind assumption of diquarks in any circumstances. Our few-body community is surprised when the three-body problem is solved by steps: (1,2) first, then (1,2)-3 as a quasi two-body problem. Similarly a four-body bound state is solved by calculating first  (1,2) and (3,4), and eventually (1,2)-(3,4), etc.  If one estimates the hydrogen molecule this way, one never gets it. Also, Borromean binding and all other subtleties of our field are lost. 
\section{Link between mesons, baryons and multiquarks}
The factor $1/2$ in \eqref{eq:H-baryons} has been discussed in several papers, in particular, Stanley and Robson \cite{Stanley:1980fe}. This rule appears naturally for one-gluon-exchange, or more generally, for any  color-octet exchange that contains a $\ll{i}{j}$ factor in front of the potential $v(r_{ij})$.  For any pair potential, the most general structure  is a combination of color-singlet and color-octet exchanges. But the former cannot contribute to the confinement, otherwise all hadrons would be stick together. So the simplest solution is to assume\footnote{For an antiquark the color generators should read $\lambda\to -\lambda^{\rm t}$}
\begin{equation}\label{eq:col-add}
 V=-\frac{3}{16}\sum_{i<j} \ll{i}{j}\,v(r_{ij})~, 
\end{equation}
for the central potential, and in particular, the potential $v(r)$ of  mesons becomes 
$\sum v(r_{ij})/2$ for baryons. This is often referred to as the ``1/2 rule''.

Amazingly, this 1/2 rule implies inequalities relating the masses, whose simplest form is 
\begin{equation}
 M(qqq)\ge \frac32\,M(q\bar q)~,
\end{equation}
which turned out to be a re-discovery of the Hall-Post inequalities invented much earlier for light nuclei. For refs., see~\cite{Richard:2019cmi}, where some extensions to tetraquarks are also given.

\begin{figure}[h!]
 \centering
 \includegraphics[width=.35\textwidth]{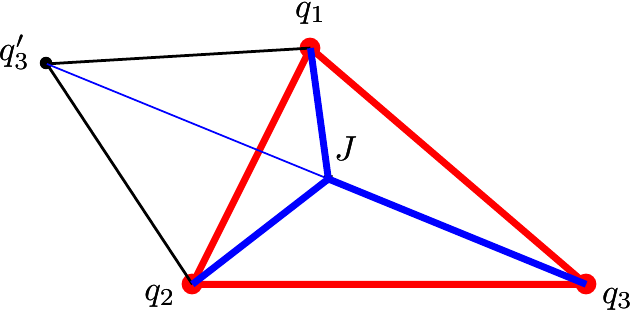}
 \caption{The length of the Fermat-Torricelli string linking the three quarks $q_1$, $q_2$, $q_3$ is equal to the distance $q_3q'_3$, where $q'_3$ forms an external equilateral triangle with $q_1$ and $q_2$. If the triangle $q_1q_2q_3$ is too flat, the length is simply  the sum of two sides. Note that there is no need to compute the location of the junction~$J$. Amazingly, $q'_3$ is also used in constructing the Napoleon triangle associated with $q_1 q_2 q_3$. }
 \label{fig:Fermat}
\end{figure}

The above discussion on the color dependence of the quark potential becomes more intricate if one gives up the restriction on pairwise interactions~\cite{Dmitrasinovic:2003cb}. 
Years ago, it was suggested that if the linear interaction $b\,r$ among a quark and an antiquark, corresponding to the string constant $b$ times the minimal path $r$ from the quark to the antiquark, it becomes for baryons the same strength $b$ times the minimal length of a three-leg string linking a central junction $Y$ to each of the quarks. This is the celebrated $Y$-shape potential shown in Fig.~\ref{fig:Fermat}. The problem of the minimal path linking three points was addressed in a correspondence between Fermat and Torricelli, and later generalized by Gauss, Steiner and many others.

As above, for the ``1/2'' rule linking mesons and baryons, and many other instances in hadron spectroscopy, the ``Mattew effect'' is at work.\footnote{See Mattew Gospel: ``So
the last will be first, and the first last'' and ``For to him who has will more be given; and from him who has not, even what he has will
be taken away'' as well as the analysis by the sociologist R. Merton~\cite{1968Sci...159...56M}}
\@
The $Y$-shape interaction was first suggested by Artru \cite{Artru:1974zn}, and then by Dosch et al.~\cite{Dosch:1975gf}, and rediscovered by many others who are often given the credit for this idea.  

For our few-body community, it is challenging to implement in three-body calculations a potential that is not pairwise. For instance, in a variational calculation, the length of the $Y$-shape should be evaluated for each set of quark positions when computing the matrix elements of the potential. 
At first, it looks necessary to compute the location of the junction $J$ and then the length $\sum Yq_i$. However, it was realized that one can calculate much more economically and directly the length of the $Y$-shape string. See, Fig.~\ref{fig:Fermat} and  \cite{Richard:2016eis} for Refs.

\begin{figure}[h!]
 \centering
 \includegraphics[width=.4\textwidth]{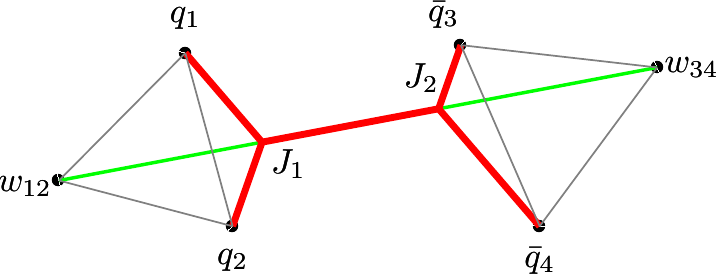}
 \caption{Planar connected string linking minimally two quarks and two antiquarks. The angles at the junctions are $120^\circ$ as in the Fermat-Torricelli string. The length of the Steiner tree drawn in red is equal to  the distance $w_{12}w_{34}$, where $w_{ij}$ makes an external equilateral triangle with the quarks $q_i$ and $q_j$. There are many particular cases occurring when three of the constituents are nearly aligned. 
 }
 \label{fig:tetra}
\end{figure}

For a tetraquark, the generalization is the double-$Y$ string, that links the two quarks to a first junction $J_1$, the two antiquarks to a junction $J_2$, and $J_1$ to $J_2$. In the planar case, it is shown in Fig.~\ref{fig:tetra}. The spatial case is more involved. The analog of the auxiliary point $C'$ belongs to the circle (named Melzak's circle) from which $q_1q_2$ is seen at $60^\circ$. There is similarly a Melzak circle for the antiquarks $\bar q_3\bar q_4$, and  the distance of interest, $d_4=q_1J_1+q_2J_1+J_1J_2+J_2\bar q_3+j_2\bar q_4$ is the maximal distance between these circles. Some tricks on how to  calculate the minimal or maximal distances between two circles in 3D can be found in articles dealing with applied geometry~\cite{5389850}, which are often used by professionals of computer-assisted cartoon drawing. Similar minimization problems in 3D are encountered for complex molecules and for proteins. 

Unfortunately, the sophisticated and appealing connected strings play a minor role if they are in competition with the \emph{flip-flop} interaction ($b$ is the string constant)
\begin{equation}
 V_{\rm ff}=b\,\min\{r_{13}+r_{24},r_{14}+r_{23}\}~,
\end{equation}
first introduced in a rather renowned paper \cite{Lenz:1985jk}, in its quadratic version. 

The string inspired potential $\min[ b\,d_4,V_{\rm ff}]$ is more attractive than the color-additive interaction \eqref{eq:col-add} applied to $v(r)=b\,r$. However, to optimize the string configuration when the quarks move, it is necessary to modify freely the color state. So the attraction induced by the string potential works really for quarks and antiquarks that are different~\cite{Vijande:2013qr}. 
\section{Multiquark spectroscopy}
\subsection{Tetraquarks}
The first multiquark configuration consists of two quarks and two antiquarks. It involves two color states, which could be chosen as ${\rm T}=\bar 3 3$ and ${\rm M}=6\bar 6$ in the $qq$-$\bar q \bar q$ basis or $11$ and $88$ in any of the $q\bar q$-$q\bar q$ bases. 

In the light sector, one of the
most celebrated analyses was based on chromomagnetism. Once the hyperfine splitting of mesons and baryons is attributed to an operator
\begin{equation}\label{eq:chromo}
 V_{ss}\propto \sum_{i=1}^N \frac{\llss{i}{j}}{m_i\,m_j}\delta^{(3)}(r_{ij})~,
\end{equation}
or its analog in the bag model, it is tempting to apply it to higher configurations. In particular, it was proposed that the lowest S-wave $qq\bar q\bar q$ could compete with the P-wave $q\bar q$ excitations in the spectrum of scalar mesons \cite{Jaffe:1976ig,Jaffe:1976ih}, the exotic $qq\bar q\bar q$ with isospin $I=2$ being pushed too high in mass to show up as narrow resonances.  

The first explicit calculation of multiquarks, to my knowledge, is in a paper by Gavela et al.\ \cite{Gavela:1978hq},devoted to $qq\bar a\bar q$, nowadays denominated ``tetraquark''. They considered a simple ``chromo-harmonic'' Hamiltonian, which after rescaling reads
\begin{equation}\label{eq:H-col}
 H=\sum_{i=1}^4 \frac{\vec p_i^2}{2}-\frac{3}{16}\sum_{i<j} \ll{i}{j}\,r_{ij}^2~,
\end{equation}
and accounted for the two possible color states, ${\rm T}=\bar 3 3$ and ${\rm M}=6\bar 6$ in the $qq$-$\bar q\bar q$ basis. If one introduces the Jacobi coordinates $\vec x=\vec r_2-\vec r_1$, $\vec y=\vec r_4-\vec r_3$ and $\vec z=(\vec r_3+\vec r_4-\vec r_1-\vec r_2)/\sqrt2$, the potential in the color states are 
\begin{equation}
\begin{aligned}
 V_{\rm TT}&= \frac34(\vec x^2+\vec y^2)+ \frac12\,\vec z^2~,\\
  V_{\rm MM}&= \frac38(\vec x^2+\vec y^2)+ \frac54\,\vec z^2~,\\
   V_{\rm TM}&= -\frac{3}{2\sqrt2}\,\vec x.\vec y~.
 \end{aligned}
\end{equation}
and the solution shows that the ground state is unstable but that some excitations are metastable, thanks to selection rules in the fall-apart decays. If one restricts to ${\rm T}$ color, the solution is analytic, and the diquark approximation consists of replacing $3/4$ by $1/2$. 

The calculation of \cite{Gavela:1978hq} was resumed, restricted to the ground state, but for  more general potentials $v(r)$ instead of $r^2$, and unequal masses \cite{Ader:1981db,Heller:1985cb}. It was found  that if the mass ratio $M/m$ is large enough, the doubly-heavy tetraquarks $QQ\bar q\bar q$ becomes bound. At first, this is purely chromo-electric effect. But it was quickly realized that if the light pair is $\bar u\bar d$ with isospin 0 and spin 0 (for the color $\bar3 3$ component), then the chromomagnetic interaction is also favorable. 

For our community who is  very interdisciplinary, one can notice the analogy between the hydrogen-like molecules $M^+M^+m^-m^-$ and the tetraquarks $QQ\bar q\bar q$. For equal masses, the positronium molecule is barely bound, and when $M/m$ increases, the molecule becomes more deeply bound and even acquires stable excitations. In the quark sector, within a central potential, the equal mass configuration is unbound, but becomes stable when the mass ratio exceeds a minimal value, under the effect of a \emph{favorable} symmetry breaking.

Why ``favorable''? For any Hamiltonian decomposed as $H=H_s+H_a$, where $H_s$ is  even and $H_a$ odd under a certain symmetry, the variational principle, with the ground state of $H_s$ as trial function, implies that the ground state decreases when $H_a$ is switched on, say $E(H)\le E(H_s)$. The problem is that the stability of a molecule or multiquark is a competition between a collective behavior and the splitting into two atoms of hadrons constituting the threshold. In most cases, breaking a symmetry benefits more to the threshold than to the collective compound, and thus tends to spoil its binding. For instance, starting from the  bound positronium molecule, which is the $M=m$ case in $M^+m^+M^-m^-$, stability disappears when $1/2.2\lesssim M/m\lesssim 2.2$~\cite{1997PhRvA..55..200B,1998NuPhA.631...91S}.

Another symmetry-breaking explains why $qq\bar q\bar q$ is unbound in the simple quark models such as \eqref{eq:H-col}, while the positronium molecule is bound. Let us start from the generic Hamiltonian
\begin{equation}
 H(\lambda)=\vec p_1^2+\cdots \vec p_4^2+ (1-2\,\lambda)\left[v(r_{12})+v(r_{34})\right]
 {}+(1+\lambda)\left[v(r_{13})+\cdots+ v(r_{24})\right]~,
\end{equation}
where $v(r)$ is attractive. Using again the variational principle, one can see that the energy is maximal for the symmetric case $\lambda=0$, and since the ground state energy is a convex function of $\lambda$, $E(\lambda')\le E(\lambda)$ if either $\lambda'<\lambda<0$ or $0<\lambda<\lambda'$.  
If one believes that the energy is nearly parabolic around $\lambda=0$, one can reasonably guess that, more generally,  $E(\lambda')\le E(\lambda)$ if $|\lambda'|>|\lambda|$. Now, for molecules, the threshold corresponds to $\lambda=-1$ with an appropriate numbering, while the Ps$_2$ configuration, with another numbering, is reached for $\lambda=2$. This explains why Ps$_2$ is stable, in a way that somehow differs from the exposition in  textbooks, but does not contradict it: the asymmetry in the coefficients of the potential explains why the $e^+e^-$ subsystems can become deformed and attract each other. For a tetraquark, the threshold still corresponds to $\lambda=2$, but the states  with color $\bar 33$ or color $6\bar 6$ are obtained for $\lambda=-1/4$ and $\lambda=7/8$, respectively. This explains why they are not bound, and it turns out that when the mixing between them is plugged in, the energy does not decrease much. For details, see \cite{Richard:2018yrm}. 

%
\subsection{Hexaquarks}
In the case of hexaquarks, there has been also several attempts to compare the 6-quark energy to the various thresholds made of two baryons. The most famed is the $H=uuddss$ configuration, for which Jaffe~\cite{Jaffe:2004ph} realized that the chromomagnetic operator 
\begin{equation}
 -\frac{3}{16} \sum_{i=1}^N \frac{\llss{i}{j}}{m_i\,m_j}~,
\end{equation}
exhibits striking coherences, with eigenvalues larger in $H$ than the cumulated values in the hadrons into which it could fall apart.  This is a rare feature. For instance in a hydrogen-like  molecule, $\sum e_i\,e_j=-2$, which coincides with $e_1\,e_3+e_2\,e_4$. So the binding of the molecule does not result from an obvious excess of attraction. In his 1977 paper, Jaffe noticed that if one adopts the limit of flavor symmetry SU(3)$_{\rm F}$, and the same short-range correlation factor $\langle \delta^{(3)}(r_{ij})\langle$ as for ordinary hadrons, then $H=uuddss$ is bound well below the baryon-baryon threshold. The same observation was made 10 years later by Gignoux et al.\ and Lipkin for the $\bar Q uuds$ pentaquark and its SU(3)$_{\rm F}$ analogs $\bar Q ddsu$ and $\bar Q ssud$.  The stability of the $H$ has been discussed in many papers, and is still debated, as some recent lattice calculations find its mass near the threshold. For refs., see,e.g., \cite{Richard:2016eis}. 

Among the effects acting against the binding of the $H$ in the quark model is the observation by Oka, Shimizu and Yazaki \cite{Oka:1983ku} that the short-range correlations are smaller in a extended dibaryon than in a compact baryon. Somewhat paradoxically, the same effect is advocated~\cite{Goldman:1989zj} in favor of the stability of the light $d^*(2380)$ dibaryon below its nominal threshold $\Delta\Delta$. In fact there is contradiction, as the chromomagnetic interaction is attractive for the $H$ and repulsive for the  $d^*(2380)$.

Recently, a lattice calculation concluded that very heavy dibaryons such as $bbbccc$ are stable against any dissociation into two baryons~\cite{Junnarkar:2019equ}. For such configurations, one could reasonably expect that a potential-model picture dominated by its chromelectric component shoulnd provide a good description. However, solving carfully the 6-body problem with standard quark models does not confirm the stability of such fully-heavy dibaryons~\cite{Richard:2020zxb}. 

Other dibaryon states have been studied, with $n<6$ units of heavy flavor, and there is no evidence for stable dibaryons \cite{Oka:2019mrd,Garcilazo:2020acl}.
\subsection{Pentaquarks}
This sector is rich of twists. In the early 60s, there has been some claim for baryon resonances with positive strangeness, but these states were nevec confirmed. See, e.g., \cite{Richard:2016eis} for refs. 
Some years later, the light pentaquark by Nakano et al;~\cite{Nakano:2003qx} was much debated, but not confirmed in high-statistics experiments. More recently, the LHCb collaboration found hidden-charm pentaquarks, with, however, some variants when the sample of data is augmented and the analysis refined~\cite{Aaij:2015tga,Aaij:2019vzc}. 

The discovery of these ``LHCb'' pentaquarks aroused a considerable interest, leading to a flurry of studies, in particular in the framework of the molecular approach, see, e.g., \cite{Guo:2015umn}. The pentaquark spectrum has also been revisited in constituent models, with the usual variants: chromomagnetism, diquarks, etc. \cite{ali2019multiquark}. In standard potential models, one cannot exclude the possibility of bound states \cite{Richard:2017una}.

Some years ago, in line with Jaffe's speculations on the $H$, it was pointed out that some $P_Q=\bar Q qqqq$ configurations benefit from the same coherences of the chromomagnetic interaction~\cite{Karl:1987uf,Richard:2019fms}. The $P_c$ state was searched for at Fermilab and at HERA, but no evidence was found~\cite{Richard:2016eis}. A full calculation of the $P_Q$, not restricted to the chromomagnetic part of the Hamiltonian, does not indicate any binding~\cite{Richard:2019fms}. 
\section{From bound states to the continuum}

Now, most experimental candidates are actually resonances, some of them being remarkably narrow. This means that the dissociation is not favored by the structure of the wave function. In the time of charmonium phenomenology, for instance, it was argued that $qq$-$\bar q\bar q$ states with an orbital momentum $\ell>0$ between the diquark and the antidiquark are protected from decaying into two $q\bar q$ mesons. 

The question is thus raised of extending the constituent models to the continuum, to separate the actual resonances from the artifacts of an approximate treatment of the states in continuum, and to estimate the widths of these resonances, if any. This concern is shared by other approaches, see, for instance, the Lüscher formulas for the lattice calculations \cite{Luscher:1986pf}.  In standard constituent models, one often uses a basis of normalizable wave functions, such as 
\begin{equation}\label{eq:var-basis}
 \Psi=\sum_{i=1}^N\gamma_i\,\exp[-(a\,\vec x^2 +2\,b\,\vec x.\vec y+\cdots)/2]~,
\end{equation}
where $\vec x,\,\vec y,\,\ldots$ are Jacobi coordinates describing the relative motion. Real scaling (or stabilization method) uses the stationarity of the wave function in the neighborhood of an eigenvalue. In particular, if one rescales one or several Jacobi variables, say $\vec x\to \alpha \,\vec x$,  the variational energy browses a plateau. If the basis becomes large, by increasing $N$ in \eqref{eq:var-basis}, there is actually a turnover of levels on each plateau with  avoided crossings. The density of states, and the minimal distance between two consecutive variational energies, give access to the width of the resonances. The method is well known for electron or positron scattering on atoms and molecules, and has been adapted to hadron physics. See, e.g., \cite{Meng:2019fan} and refs.\ there. 

Another strategy is complex scaling, which transforms the oscillatory behavior of resonances into a decreasing one. A review can be found, e.g., in \cite{Lazauskas:2019ltg}. The method can be combined with a variational expansion such as~\eqref{eq:var-basis} whose range coefficients become complex. Again, the stationarity of the trial energy is a signature of resonances. Some interesting benchmark cases have been treated by Okopinska et al.\ \cite{2013JPhA...46h5303K}, with an expansion into the eigenstates of a single oscillator rather than on a basis of Gaussians. The same strategy could perhaps be applied to multiquark resonances in quark models.  
\vskip .2cm
\section{Outlook}
Contemporary hadron physics involves very elaborate techniques such as QCD sum rules or lattice simulations. Even the so-called ``simple'' constituents models require a careful treatment of the few-body problem. It is often read that baryon = quark + diquark, and tetraquark = diquark + antidiquark, with the consequence of deforming the predictions of the model. Actually, a careful treatment of the $QQ'\bar q\bar q'$ problem gives a very rich palette of effects: meson-meson configurations at large distances, diquark and/or antidiquark clustering in some regions,  color-sextet or color-octet excitations for some pairs. The bound state part of the spectrum can be calculated in a rather straightforward way, e.g., by standard variational methods, though it is often treated with some unjustified simplifications.
The most salient feature is the scarcity of bound states, restricted to very peculiar configurations. 
Novel possibilities might show up for multiquark with different quarks, in which the color fluxes could rearrange freely without any constraint from antisymmetrization. 

The spectrum of resonances requires dedicated techniques, which are necessary to compare the results with the rich set of experimental data. 
\subsection*{Acknowledgments}
Along the years, I benefited from several informative and friendly discussions with the late Mikhail Voloshin. I~also thanks M.~Asghar for comments on the manuscript. 
%
%

%
\end{document}